\documentclass[letterpaper,11pt]{article}
\usepackage{geometry} %[margin=1in]
\usepackage[utf8]{inputenc}
\usepackage[T1]{fontenc}
\usepackage{lmodern}

\usepackage[dvipsnames,table]{xcolor}
\usepackage{graphicx}
\graphicspath{{./}{figures/}}
\usepackage{hyperref} %[colorlinks]
\hypersetup{colorlinks = true, 
            linkcolor=black,
            urlcolor=RoyalBlue,
            citecolor=black,
            }
\usepackage{url}
\usepackage{gensymb}
\usepackage{fancyhdr}
\usepackage{wrapfig}
\usepackage{sidecap}
\usepackage{natbib}  %[square,numbers]
\usepackage{amsmath}
\usepackage[nameinlink,noabbrev]{cleveref}
\usepackage{amssymb}
\usepackage{tcolorbox}
\usepackage{longtable}
\usepackage{enumitem}
\usepackage{floatrow}
\newfloatcommand{capbtabbox}{table}[][\FBwidth]
\usepackage{lipsum}
\usepackage[displaymath, mathlines]{lineno}
\modulolinenumbers[2]
\usepackage{float}
\usepackage{pdfpages} %to use \includepdf
\usepackage{todonotes}
\usepackage{sectsty}
\usepackage{authblk}

\sectionfont{\fontsize{12}{15}\selectfont}
\subsectionfont{\fontsize{12}{15}\selectfont}

\title{\vspace{-3em}Field Locations}

\definecolor{lightgray}{gray}{0.9}
\date{\vspace{-5ex}}

\begin{document}
% \maketitle

\noindent
\textbf{\large Roman CCS White Paper}

\vspace{2em}
\begin{center}
{\Large Considerations for Selecting Fields for the Roman High-latitude\\\vspace{0.2em}Time Domain Core Community Survey}
\end{center}

\vspace{1.5em}
\noindent
\textbf{Roman Core Community Survey:} High Latitude Time Domain Survey

\vspace{0.5em}
\noindent
%Choose from https://hst-docs.stsci.edu/hsp/hubble-space-telescope-call-for-proposals-for-cycle-31/hst-filling-out-the-apt-phase-i-proposal-form
\textbf{Scientific Categories:} stellar physics and stellar types; stellar populations and the interstellar medium; large scale structure of the universe
% TDAMM

\vspace{0.5em}
\noindent
%suggestions for each can be found https://hst-docs.stsci.edu/hsp/hubble-space-telescope-call-for-proposals-for-cycle-31/appendix-b-scientific-keywords
\textbf{Additional scientific keywords:} Supernovae, Cosmology, Dark energy

\vspace{1em}
\noindent
\textbf{Submitting Author:}\\
Benjamin Rose, Baylor University (Ben\_Rose@baylor.edu)\\

\vspace{0.5em}
\noindent
\textbf{List of contributing authors:}\\
% Name, affil (email)\\
Greg Aldering, Lawrence Berkeley National Lab (galdering@lbl.gov)\\
Rebekah Hounsell, University of Maryland Baltimore County, NASA Goddard Space Flight Center 
(rebekah.a.hounsell@nasa.gov)\\
Bhavin Joshi, Johns Hopkins University (bjoshi5@jhu.edu)\\
David Rubin, Univserity of Hawaii (drubin@hawaii.edu)\\
Dan Scolnic, Duke University (dan.scolnic@duke.edu)\\
Saul Perlmutter, University of California, Berkeley (saul@lbl.gov)\\
Susana Deustua, NIST (susana.deustua@nist.gov)\\
Masao Sako, University of Pennsylvania (masao@sas.upenn.edu)

\vspace{1em}
\noindent
\textbf{Abstract:} \\
In this white paper, we review five top considerations for selecting locations of the fields of the Roman High-latitude Time Domain Survey.  
% Based on these considerations, 
% we recommend specific fields to study, and propose that these fields are selected well-ahead of the mission start to maximize community synergy. \textbf{TODO: write an abstract}
Based on these considerations, we recommend Akari Deep Field South (ADFS)/Euclid Deep Field South (EDFS) in the Southern Hemisphere has it avoids bright stars, has minimal Milky Way dust, is in Roman Continuous viewing zone, overlaps with multiple past and future surveys, and minimal zodiacal background variation. In the North,  Extended Groth Strip (EGS) is good except for its zodiacal variation and Supernova/Acceleration Probe North (SNAP-N) and European Large Area Infrared Space Observatory Survey-North 1 (ELAIS N-1) are good except for their synergistic archival data.
\thispagestyle{empty}
\newpage
\setcounter{page}{1}

% \url{https://roman.gsfc.nasa.gov/science/ccs_white_papers.html}
% The suggested length is 2 or 3 pages of text, plus figures, tables, and references as needed.  

% \linenumbers
\section{Introduction}

% \textbf{Todo: make sure all acronyms are defined at the right time.}

The choice of field location for the Roman's High Latitude Time Domain Survey is one of the most critical ones for planning the survey, and a decision that well precedes the onset of the survey allows the community to prepare for synergies across the electromagnetic spectrum.  Here, we review the key considerations of field selections, and make recommendations based on the constraints.  Other white papers will discuss the number and size of the survey fields, here we will focus solely on the trades in field locations.  The focus of this particular white paper is led by optimization of a cosmological survey with Type Ia Supernovae (SNe~Ia), but we note many other astrophysical studies are impacted by the choice of this field.

We quantify five key selection criteria when choosing the location of the deep fields.
\begin{enumerate}
    \setlength\itemsep{0em}
    \item Avoid bright stars
    \item High Galactic latitude to minimize dust extinction and MW stars
    \item High ecliptic latitude to minimize zodiacal light, and to reach the Roman Space Telescope continuous-viewing zone  to avoid observational seasons and meet Roman science requirement SN 2.3.4
    \item Overlap with past, current, and planned wide-area surveys
    % \item Avoid LMC and SMC dust and stars
    \item Seasonal variation of zodiacal background
\end{enumerate}

\subsection{Avoiding bright stars}
\begin{figure}[t]
    \centering
    \includegraphics[width=0.8\textwidth]{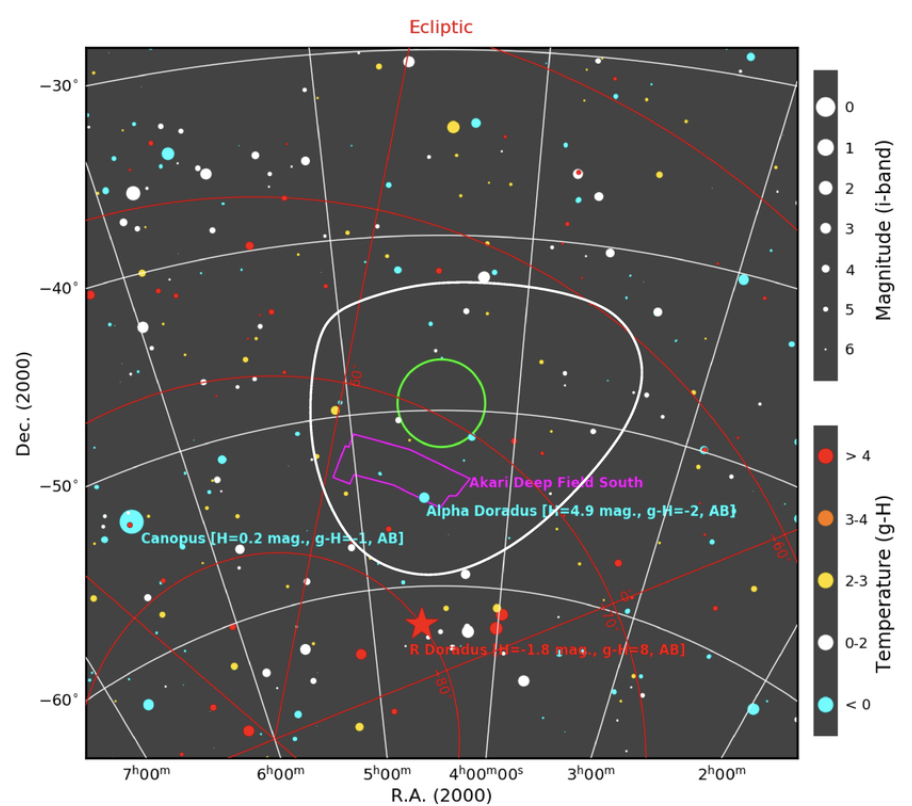}
    \caption{Created for the Euclid Survey while studying deep field locations (credit to Jean-Charles Cuillandre). The positioning of the Euclid Deep Field South (green) and Akari Deep Field South (purple) relative to bright stars in the area.}
    \label{fig:euclid}
\end{figure}

This consideration is often overlooked, but recently necessitated a change in the position of Euclid's Deep Field \footnote{\url{https://sci.esa.int/web/euclid/-/61403-three-dark-fields-for-euclid-deep-survey}}.  The magnitude of a bright-star cutoff must be studied; for Euclid, this was around 6$^{th}$ magnitude in the I band.  Due to the higher sensitivity of Roman, it may be advantageous to place a cutoff closer to the 7$^{th}$ magnitude.  However, it is unclear how feasible it may be to avoid fainter stars than 6$^{th}$  simply due to the number of them across the sky.  A figure from the Euclid study is shown as \cref{fig:euclid}.

\begin{figure}[t]
    \centering
    \includegraphics[width=0.8\textwidth]{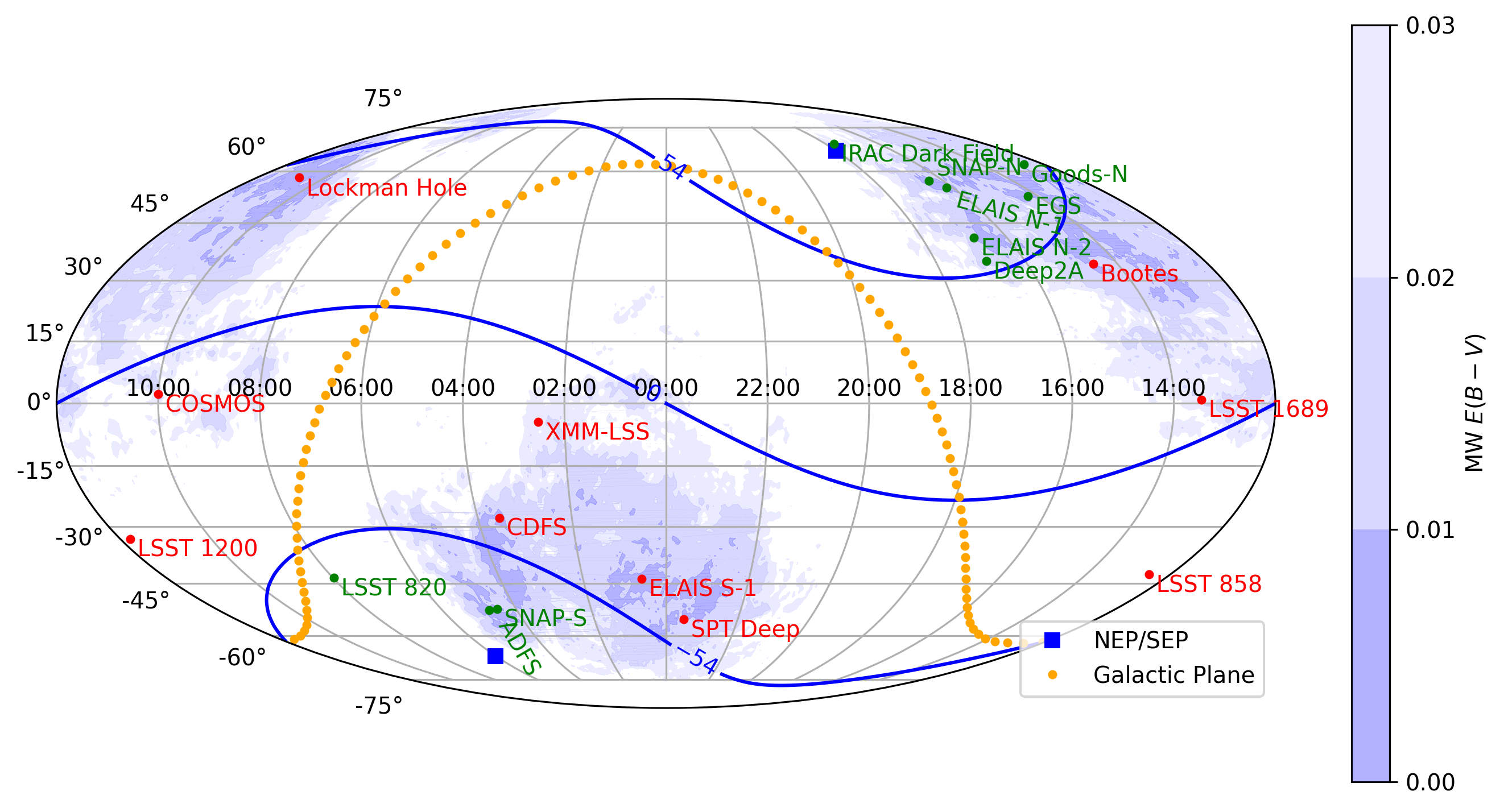}
    \caption{An equatorial all-sky plot of possible Roman time-domain fields. Green labels are in the current Roman CVZ ($\pm54^{\degree}$ off the ecliptic) and red labels are outside the CVZ. Low Milky Way dust extinction is shown as blue shading. Overall, the top field choices include Extended Groth Strip (EGS), and Akari Deep Field South/Euclid Deep Field South (ADFS). We note that if the field of regard is improved by $\sim$8$^{\degree}$, some particularly attractive fields such as the Chandra Deep Field-South (CDFS) would become accessible.}
    \label{fig:skymap}
\end{figure}

\subsection{Minimizing Milky Way dust extinction}

While Milky Way (MW) extinction must be considered for maximizing the depth of the observations, it is also a key systematic for studies of SN~Ia cosmology. MW extinction uncertainty is fundamentally different from host-galaxy extinction because, while the latter imprints on the rest frame, MW extinction imprints on the observer frame. This translates into different amounts of correction in the rest frame, i.e., as a function of redshift, and thus can have covariance with the cosmological parameters. At lower values of MW extinction the absolute spatial variance in the extinction is also smaller which helps to reduce effect.

The $A_V$ values of various fields are given in Table~1.  We recommend a maximum $A_V$ per field of $0.04$ to limit the impact of this systematic uncertainty. For low MW dust, we prefer Supernova/Acceleration Probe North (SNAP-N), European Large Area Infrared Space Observatory Survey-North 1 (ELAIS N-1), Extended Groth Strip (EGS), Akari Deep Field South (ADFS)/Euclid Deep Field South (EDFS)\footnote{\url{https://www.cosmos.esa.int/web/euclid/euclid-survey}}, Supernova/Acceleration Probe (SNAP-S), and the JWST North Ecliptic Pole Time Domain Field (JWST NEP TDF). % ; \citealt{Jansen2018}). 

% We like: CDFS\footnote{CDFS will have gap on the sun-ward side even if the CVS is extended anti-sun}, EGS, SPT Deep, ADFS

\begin{table}[t]
\begin{center}
{Table 1: Possible CVZ fields for the\\Roman High-latitude Time Domain Survey}
\vskip 5pt
\begin{tabular}{lccc}
\hline \\[-0.8em]
Name & \multicolumn{1}{c}{RA} & \multicolumn{1}{c}{Dec}& $A_V$ \\
   & (deg) & (deg) & (mag)\\
\hline\hline \\[-0.7em]
\hspace{-1em}\textit{North Hemisphere}\\
SNAP-N          &  246.25 & $  +57.0$ &  0.020\\
ELAIS N-1       &  242.75 & $  +55.0$ &  0.021\\
EGS             &  214.25 & $  +52.5$ &  0.022\\
JWST NEP TDF    &  260.70 & $  +65.8$ &  0.030\\
Goods-N         &  189.19 & $  +62.2$ &  0.032\\
ELAIS N-2       &  251.70 & $  +41.0$ &  0.040\\
Deep2A          &  253.00 & $  +34.9$ &  0.048\\
IRAC Dark Field &  265.00 & $  +69.0$ &  0.120\\ 
\hline
\hspace{-1em}\textit{Southern Hemisphere}\\
ADFS/EDFS       &   71.00 & $  -52.3$ &  0.021\\
SNAP-S          &   67.50 & $  -52.0$ &  0.022\\
LSST 820        &  119.56 & $  -43.4$ &  0.940\\
\hline
% \multicolumn{4}{l}{\hspace{-2em}$A_V$ values are for \citet{Schlafly2011} via}\\
% \multicolumn{4}{l}{\hspace{-2em}\url{https://irsa.ipac.caltech.edu/applications/DUST/}.}
\end{tabular}
\label{table}
\\\vspace{0.4em}
$A_V$ values are for \citet{Schlafly2011} via \url{https://irsa.ipac.caltech.edu/applications/DUST/}.
\end{center}
\end{table}

\subsection{Continuous Viewing Zone}

Seasonal gaps in coverage are sub-optimal, especially for high redshift transients, a unique capability of Roman.  The typical light curve length at rest frame for a SN~Ia is $\sim45$ days.  At a redshift of $z=1.5$, this would be almost 4 months. Therefore, edge effects on the sample due to seasonality can be quite serious.

The obvious recommendation is for the fields to be in the Continuous Viewing Zone (CVZ, $>$54$^{\degree}$ above/below the ecliptic).  The Declination limits for the CVS vary by RA. The CVZ limit ranges from  $\pm$77$^{\degree}$ and $\pm$31$^{\degree}$. \Cref{fig:skymap} in an all sky plot showing the CVZ, the Galactic plane, Milky Way dust, and candidate fields. 

\subsection{Synergies with other surveys \& instruments}

Overlap with other surveys is important for a number of reasons.  This includes observations in different parts of the electromagnetic spectrum (e.g., UV/optical), spectroscopic follow-up of transients and acquisition of host-galaxy redshifts. For different fields, there are a lot of different avenues for synergies with past and current surveys. For a detailed list of synergies specifically between the largest surveys of the next decade---Roman, Rubin Observatory's LSST, and Euclid---see \citealt{Rose2021b}.

There are many fields with rich archival data as well as future Rubin, Euclid and DESI data, such as Chandra Deep Field-South (CDFS), that are not in the CVZ. CDFS is the closest, just 8$^{\degree}$ outside. Though ADFS/EDFS will also be observed with Rubin and Euclid and also has archival data. 

% In the north, Northern visible:
Roman has in-kind time on Subaru PFS, the largest multi object spectrograph of the 2030s. Since Subaru is in the Hawaii, it can not observe Roman's Southern CVZ. With extremely high airmass, Subaru can observe down to a declination of $\sim-$30$^{\degree}$, the declination of CDFS. DESI can also obtain redshifts down to $m_{AB}\sim25$ in the northern sky.

% Added by BAJ
Another field to consider with rich pre-existing archival data is the JWST NEP TDF \citep{Jansen2018}. Overlap of the HLTDS deep tier with the JWST NEP TDF could be beneficial given the extant data from X-ray (Chandra; $\lesssim$1~keV) to radio (LOFAR; 150~MHz) along with significant investment of time from HST, JWST, and many ground-based observatories. This field of roughly 14' diameter was selected to be free of foreground stars brighter than $\sim$15.5 mag, low MW extinction ($A_V \sim 0.03$), and in the northern CVZ.

% In the south, Spitzer, the Dark Energy Survey, the Vera C. Rubin Observatory, Euclid
Of the northern fields with minimal Milky Way dust---SNAP-N, ELAIS N-1, JWST NEP TDF, and EGS---JWST NEP TDF and EGS have the broadest archival data set.

% Roman high-latitude wide area ..
The Roman High-latitude Time Domain survey should also consider synergies with the Roman High-latitude Wide Area Survey. The Wide Area Survey requires a deep spectroscopic field. If this field overlapped with the Time Domain field, then the same galaxy redshifts can be used for multiple core Roman science cases.

\subsection{Seasonal variation of zodiacal background}

High ecliptic latitude reduces zodiacal background as well was keeps a field in the Roman CVZ. When looking at the zodiacal background and variability throughout the year, using an ecliptic latitude of $<$15$^{\degree}$ from the ecliptic poles is a good proxy for minimizing seasonal variation.
When performing a full calculation (by using \texttt{ZodiPy}, \citealt{San2022}), EGS has about twice the seasonal variation compared to SNAP-N, ELAIS N-1, and ADFS/EDFS.

% Here we use the ZodiPy [San
% et al., 2022] to compare the survey speed and homogeneity of these fields within the RST continuous
% viewing zone based on their zodiacal light level compared to the darkest good field (the IRAC Dark
% Field) and the fractional range in that level over the course of a year. Table 1 shows the results. After
% the first five fields the survey speed for a prism survey drops quickly and the inhomogeneity increases
% substantial (i.e., the first 5 fields have inhomogeneity around 20% across the year, but then it rises to
% 40% or much more for fields with lower speeds). Ecliptic latitude (ε) also serves as a pretty good guide,
% but again, above |ε| > 75◦, Galactic extinction and stellar interference become worse

\section{Conclusion}

After going through five main selection criteria, we present possible fields, \cref{table}. Of these possible fields, ADFS/EDFS in the Southern Hemisphere has minimal Milky Way dust, is in the CVZ, overlaps with multiple past and future surveys, and minimal zodiacal variation. In the North, EGS is good except for its zodiacal variation and SNAP-N and ELAIS N-1 are good except for their synergistic archival data.

\bibliographystyle{apj}
\bibliography{library}

\end{document}